\documentstyle[12pt]{article}

\newcommand{\beq}{\begin{equation}}
\newcommand{\eeq}[1]{\label{#1}\end{equation}}
\newcommand{\beqn}{\begin{eqnarray}}
\newcommand{\eeqn}[1]{\label{#1}\end{eqnarray}}

\newcommand{\ds}{\!\!\not\!\partial}
\newcommand{\as}{\not\!\! a}

\newcommand{\D}{{\cal{D}}}

\newcommand{\Z}{{\cal Z}}

\newcommand{\Las}{{\cal L}}
\newcommand{\bp}{\psi^{\dagger}}

\begin{document}
\begin{titlepage}
\setcounter{footnote}0
\begin{center}
{\LARGE\bf Spinons as composite fermions}\\[.4in]
{\Large Daniel C. Cabra\footnote{CONICET, Argentina.
E-mail address: cabra@venus.fisica.unlp.edu.ar} and
Gerardo L. Rossini\footnote{CONICET, Argentina.
E-mail address: rossini@venus.fisica.unlp.edu.ar}}\\
\bigskip {\it Departamento de F\'\i sica,  \\
Universidad Nacional de La Plata,\\
C.C. 67 (1900) La Plata, Argentina.}\bigskip\\
\end{center}
\bigskip

\centerline{\bf ABSTRACT}
\begin{quotation}
We show that gauge invariant composites in the fermionic realization of
$SU(N)_1$ conformal field theory explicitly exhibit the holomorphic
factorization of the corresponding WZW primaries. In the $SU(2)_1$ case
we show that the holomorphic sector realizes the spinon $Y(sl_2)$             
algebra, thus allowing the classification of the chiral Fock space in
terms of semionic quasi-particle excitations created by the composite
fermions.
\end{quotation}
\end{titlepage}


\noindent {\it i) Introduction}

\vspace{.5cm}

\setcounter{footnote}{0}
The notion of generalized statistics has been the
subject of active research in the last ten years. Firstly, it was
proven that
particles obeying fractional statistics, intermediate between bosons and
fermions, can exist in $2+1$ dimensions \cite{Wil}. Then, the
possibility that also $1+1$ dimensional systems admit fractional
statistics excitations was recognized (in this case the
notion of generalized statistics must be
introduced) \cite{Hal-spinon,Hal-stat}.
Now, in spite of the amount of effort devoted to the subject, a general
formulation of quantum statistical
mechanics for systems whose excitations obey fractional statistics, as
spinons in $1+1$  or anyons in $2+1$ dimensions, has not
been fully developed, although progress in this direction  has been made
in Refs. \cite{Hal-stat,Wu}.

Concerning the physical interest of these matters,
a physical manifestation of $1+1$ fractional statistics behavior in
continuous systems has arised, for instance, in the study of the
incompressible quantum
Hall liquid. It appears in connection with the two-dimensional
modes in the boundary whose dynamics is governed by a Conformal Field
Theory (CFT) \cite{Wen}.
The study of field theories whose excitations obey generalized
statistics has also received much attention in connection with the
so-called spinon representation of CFT
\cite{Hal-spinon,Hal-al,Bernard,Ludwig}.

For the reasons above, it is important to
have as much information as possible concerning  $1+1$ dimensional systems
manifesting these properties, particularly that
arising from an explicit Lagrangian
formulation (which is lacking).
A thorough comprehension of the system properties, including its
statistical and thermodynamical behavior, could be better achieved in
this context.

The purpose of the present note is twofold:

i) We show that the $SU(N)_1$ WZW primary fields can be {\it explicitly} 
factorized into holomorphic and anti-holomorphic fields (a fact that is 
usually heuristically assumed). The holomorphic factors correspond to the 
simplest BRST invariant fields in a fermionic coset representation of the 
model.

ii) In connection with the recently revealed Yangian structure
of these theories \cite{Hal-al,Bernard,Ludwig,Sch}
we show that the BRST invariant fields mentioned above exactly
behave as the spinon fields of Refs. \cite{Bernard,Ludwig}. In this way we
provide a fermionic {\it Lagrangian} formulation for spinons, in the form of a
well-known two-dimensional field theory in which the physical (BRST invariant)
operators naturally exhibit $1+1$ generalized statistics
\cite{Hal-stat}.

We start by studying the $SU(N)_1$-WZW CFT formulated
as a fermionic coset model. This representation
provides an explicit holomorphic factorization of the primary
fields, in which the holomorphic factors are indeed given by the BRST 
invariant composite fermions.

As discussed in Refs. \cite{Hal-al,Bernard,Ludwig,Sch}, there exists an
underlying Yangian $Y(sl_N)$ symmetry in the $SU(N)_1$-WZW CFT. This
symmetry ensures that the whole Fock space of the theory can be
decomposed into multiplets which are irreducible representations of
$Y(sl_N)$.
In the $N=2$ case this construction is extensively developed and 
it was proven that the
excitations created by the so called spinon fields span the
whole chiral Fock space of the theory. The excitations created by
these spinon fields have shown up to behave as semionic quasiparticles.

We show that the holomorphic factors that we present
for the spin 1/2 primary field in the $SU(2)_1$ CFT exactly behave
as the spinon fields of Refs. \cite{Bernard,Ludwig}, thus providing a
fermionic Lagrangian formulation for spinons.
In this respect, our construction should be compared with the original bosonic
formulation, where a
heuristic holomorphic factorization of the WZW fields is assumed and no 
Lagrangian is given, but
the dynamics is instead recovered from the conformal properties of the fields.

This letter is organized as follows: in section $ii)$ we prove the
holomorphic factorization of the fundamental representation $SU(N)_1$
primary fields in terms of
gauge invariant fermions. In section $iii)$ we discuss the fermionic
(Lagrangian) representation of $SU(2)_1$ spinon fields. In section $iv)$
we present the factorization of the rest of the primaries in $SU(N)_1$
and 
comment on their relation with the Yangian symmetry. Finally, we
summarize our results in section $v)$. For the sake of conciseness we
also recall the fundamentals of the Yangian algebra in the appendix.

\vspace{.5cm}

\noindent {\it ii) Holomorphic factorization of WZW fields}

\vspace{.5cm}

It is known that the $SU(N)_1$ CFT can be formulated as a
constrained fermionic model,
that is as a $U(N)/U(1)$ fermionic coset theory \cite{NS}.
The constraint is imposed on a system of $N$ free Dirac fermions by
requiring that physical states $|phys\rangle$ are singlet under the $U(1)$
current,
\beq
J_{\mu}|phys\rangle=0.
\eeq{3.a}
This is achieved in the path integral formulation by introducing a
Lagrange multiplier $a_{\mu}$ which acts as a $U(1)$ gauge
field with no dynamics.
The partition function for the constrained model in Euclidean space
reads
\beq
\Z=\int \D\bp \D\psi \D a_{\mu} \exp(-\int \Las d^2x),
\eeq{3.b}
where the fermionic Lagrangian $\Las$
is given by
\beq
\Las=\frac{1}{\sqrt{2}\pi}{\bp}^i(\ds +i\as)\psi^i .
\eeq{3.c}
Here $\psi^i$, $i=1,\dots,N$, are Dirac fermions and $a_{\mu}$ is the
abelian gauge field implementing the constraint.
This procedure removes the charge degree of freedom from the fermionic
system.

The theory described by the Lagrangian
(\ref{3.c}) has a local  $U(1)$ invariance, thus its physical operators
will be those invariant under $U(1)$ gauge transformations.
In the present case they are constructed as \cite{CR1}
\beq
\begin{array}{l}
\hat{\psi}^i=\exp(-i\int_x^{\infty}a_{\mu}dx_{\mu})\psi^i,\\
\hat{\psi}^{i \dagger}=
{\psi^{i \dagger}}
\exp(i\int_x^{\infty}a_{\mu}dx_{\mu}).
\end{array}
\eeq{3.2}
These fields are independent of the path chosen for the integration in
the exponential due to the $a_{\mu}$ equations of motion.

It is convenient now to pass to complex coordinates
$z=\frac{1}{\sqrt 2}(x_0+ix_1)$,
$\bar{z}=\frac{1}{\sqrt 2}(x_0-ix_1)$ and to write the partition
function in a decoupled form. This is achieved
through the following change of variables
\beq
\begin{array}{ll}
a=i(\bar{\partial}u)u^{-1} &
\bar{a}= i(\partial\bar{u})\bar{u}^{-1},  \\
\psi_1=u\chi_1 & \bp_2=\chi_2^{\dagger} u^{-1},\\
\psi_2=\bar{u}\chi_2 & \bp_1=\chi_1^{\dagger} \bar{u}^{-1}.
\end{array}
\eeq{3.d}
where
$\partial \equiv \frac{\partial}{\partial z}$,
$\bar{\partial} \equiv \frac{\partial}{\partial \bar z}$
and $\psi=\left(
\begin{array}{c}
\psi_1 \\ \psi_2
\end{array}
\right)$.
The fields $u$ and $\bar u$ are parametrized in terms of real scalar
fields as
$u=\exp(-\phi -i\eta)$ and
$\bar u=\exp(\phi -i\eta)$.

Taking into account the gauge fixing procedure and the Jacobians
associated to (\ref{3.d}) \cite{FiPol} one arrives at
the desired decoupled form for the partition function (\ref{3.b}):
\beq
\Z=\Z_{ff}\Z_{fb}\Z_{gh},
\eeq{3.e}
where
\beqn
\Z_{ff} &=& \int \D\chi^{\dagger} \D\chi \exp( -\frac{1}{\pi}
\int(\chi^{\dagger}_2 \bar{\partial} \chi_1 +
\chi^{\dagger}_1 \partial \chi_2)d^2x), \nonumber \\
\Z_{fb} &=& \int \D\phi \exp(\frac{N}{2\pi}\int \phi \Delta \phi d^2x),
\nonumber
\\
\Z_{gh} &=& \int \D\bar{c} \D c \D \bar{b} \D b \exp(-\int(b\partial c +
\bar{b}\bar{\partial} \bar{c} )d^2x ).
\eeqn{3.e'}
Notice that, although the partition function of the theory is completely
decoupled, BRST quantization conditions connect the different sectors in 
order to ensure unitarity \cite{Karabali}.
Once we have obtained the desired decoupled expression for the partition
function of the $SU(N)_1$ model we proceed further to study its spectrum.

In the decoupled picture, the components of the gauge invariant  fields
(\ref{3.2}) can be written as
\beq
\begin{array}{ll}
\hat{\psi_1}^i(z)
=e^{-\varphi(z)}\chi_1^i(z)
& \hat{\psi_1}^{i \dagger}(\bar z) =
\chi_1^{i \dagger}(\bar z) e^{-\bar{\varphi}(\bar z)}
\\
\hat{\psi}_2^i(\bar z) =
e^{\bar{\varphi}(\bar z) }\chi_2^i(\bar z)
& \hat{\psi_2}^{i \dagger}(z)
=\chi_2^{i \dagger}(z) e^{\varphi(z)}
\end{array}
\eeq{3.3}
where
\beq
\begin{array}{l}
\varphi (z)= \phi+i\int_x^{\infty} dz_{\mu} \epsilon_{\mu\nu}
\partial_{\nu}\phi \\
\bar{\varphi} (\bar z)= \phi-i\int_x^{\infty} dz_{\mu} \epsilon_{\mu\nu}
\partial_{\nu}\phi
\end{array}
\eeq{3.4}
are the chiral (holomorphic and anti-holomorphic) components of the free
boson $\phi$, satisfying
\beq
\bar{\partial} \varphi=\partial \bar{\varphi}=0.
\eeq{3.f}
This fact together with the equation of motion of the free fermions
$\chi$ ensures
that $\hat{\psi}^i_1$ and $\hat{\psi}^{i\dagger}_2$ ($\hat{\psi}^i_2$
and
$\hat{\psi}^{i\dagger}_1$)
are holomorphic (anti-holomorphic).

Eq.(\ref{3.3}) makes evident the relevance of the gauge invariance
requirement for physical operators: physical excitations created by
gauge invariant composites $\hat{\psi}_1^i$, $\hat{\psi}_2^{i\dagger}$
are automatically holomorphic,
while those created by $\hat{\psi}_2^i$, $\hat{\psi}_1^{i\dagger}$  are
anti-holomorphic .

Since the composite fermions in (\ref{3.2}) are now completely given
in terms of free fields, their
properties can be easily studied. In particular, we
are interested in their short distance operator product expansion (OPE).
Being the composite fermions primary fields, the leading singularity in
their OPE is governed by their conformal dimensions $h$ and $\bar h$.
This in turn
is simply given by the sum of the dimensions of
their decoupled constituents (see eqs.(\ref{3.3}) and (\ref{3.4})).
For the holomorphic composites $\hat{\psi}_1^i$ and
$\hat{\psi}_2^{i\dagger}$ one has

\beq
h=\frac{1}{2}-\frac{1}{2N}=\frac{N-1}{2N}, ~~~~\bar{h} =0.
\eeq{3.5}
(Similarly, $h=0$ and $\bar{h} =\frac{N-1}{2N}$ for the anti-holomorphic
composites  $\hat{\psi}_2^i$ and $\hat{\psi}_1^{i\dagger}$).

This altogether gives place to the following OPE:
\beq
\hat{\psi}^{i\dagger}_2(z)
\hat{\psi}^j_1(w)=\frac{\delta^{ij}}{(z-w)^{(N-1)/N}}+\cdots
\eeq{3.6}
(Analogously
$ \hat{\psi}^{i\dagger}_1(\bar z)
\hat{\psi}^j_2(\bar w)
=\frac{\delta^{ij}}{(\bar z-\bar w)^{(N-1)/N}}+\cdots $).

These composite fields can indeed be seen as the holomorphic factors of
the WZW
primaries $g$, $g^{\dagger}$ in the fundamental representation of
$SU(N)$.
This fact follows from the consideration of the bosonization dictionary
\cite{NS} and the non observability of phase factors in fermion
bilinears, which allow us to write
\beq
g^{ij}(z, \bar z)=
\psi_2^i\psi_2^{j\dagger}=
{\hat{\psi}}_2^i(\bar z){\hat{\psi}}_2^{j\dagger}(z)
\eeq{3.7}
This equation provides
the explicit holomorphic factorization of the WZW primary field
$g$. In the same way one gets

\beq
(g^{\dagger})^{ij}(z, \bar z)=
\psi_1^i\psi_1^{j \dagger}=
{\hat{\psi}}_1^i(z){\hat{\psi}}_1^{j \dagger}(\bar z).
\eeq{3.8}
Eqs. (\ref{3.7}) and (\ref{3.8}) define the holomorphic factors of $g$
and $g^{\dagger}$ up to an arbitrary constant phase.

Let us recall that the factors $\hat{\psi}$ and ${\hat{\psi}}^{\dagger}$
independently create physical
excitations in contrast with the original fermions which do not.
This result is one of the main points in our work, that is the
explicit factorization of the $g$ WZW primary field in terms
of holomorphic and anti holomorphic gauge invariant fermions.

\vspace{5mm}

\noindent {\it iii) $SU(2)_1$ spinons}

\vspace{5mm}

We will now show that the holomorphic
gauge-invariant composite fermions
in the fermionic coset formulation of the $SU(2)_1$ CFT do fulfill the
characterization of the holomorphic part of the spin 1/2 primary field.
In particular, we show that they provide a Lagrangian construction for
spinon fields.

The spinon formulation of $SU(2)_1$ WZW theory has been recently
developed \cite{Bernard,Ludwig}. It exploits the representation theory
of the Yangian $Y(sl_2)$ algebra (see the appendix), a symmetry present
in the model, in order to give a new representation of its chiral
spectrum. The spectrum thus obtained is naturally organized into Yangian
multiplets, each one associated to a Yangian highest weight state, and
turns out to be exactly equivalent
to the standard construction based on the representation theory of
Virasoro and Kac-Moody algebras.

One can construct the Yangian highest weight states in the
context the $SU(2)_1$ WZW theory
\cite{Hal-al,Ludwig}.
Moreover, in these references each state in the spinon 
Fock space is constructed by the repeated
action of the Laurent modes of a spinon field on the vacuum (we
summarize this construction below), where the spinon field is defined as
the ``holomorphic part'' of the spin 1/2 Virasoro primary field.
Rigorously speaking, one assumes a heuristic
holomorphic factorization of the
WZW primary fields $g(z, \bar{z})=\phi (z)\bar{\phi}(\bar{z})$ and then
makes use of OPEs that follow from the action of Virasoro and Kac-Moody
generators on primary fields, as was done in Ref. \cite{Ludwig}.
An alternative construction was given in Ref. \cite{Bernard} in
terms of chiral vertex operators motivated by the
bosonization of two complex Dirac fermions with an $SU(2)$ charge degree of
freedom arbitrarily wiped out.

In our construction below the holomorphic
factors directly follow from the fermionic coset Lagrangian (\ref{3.c})
and from the principle of BRST invariance for physical operators. Also
their OPEs can be derived from the partition function by standard field
theoretical methods. The charge degree of freedom of the original
fermions is also naturally removed from the theory by the coset
constraining procedure, in a situation which is reminiscent
of what happens in the original scenario for spinons, namely the
Haldane-Shastry
spin chain, where the charge degree of freedom is
fixed by construction \cite{Hal-old,Shastry}.

In order to make contact with the spinon formulation discussed in
Ref.\cite{Ludwig} we are going to first study some
details of the $SU(2)_1$ case. In particular,
the $SU(2)$ fundamental representation and its conjugate are equivalent,
so the matrix elements of the WZW field $g$ and its dagger $g^{\dagger}$
are related by
\beq
g^{ij}=-\epsilon^{jl}(g^{\dagger})^{lk}\epsilon^{ki}.
\eeq{3.9}
This relation does not trivially apply to the fermi bilinears
(\ref{3.7})
and (\ref{3.8}), since it leads to inconsistencies in correlators.
As an example, consider the two-point correlator \cite{KZ,FZ}
\beq
\langle
g^{ij} (g^{\dagger})^{kl}
\rangle
=-\epsilon^{kr}\langle g^{ij} g^{rs}\rangle \epsilon^{sl}
\neq 0   .
\eeq{3.9'}
The inconsistency is apparent when one uses the bosonization rules to
evaluate  $\langle g^{ij} g^{rs} \rangle $ in terms of fermions, thus
obtaining a vanishing result.
Notice that a similar situation is already present in Witten's
original bosonization rules for the $U(N)$ case \cite{Witten}.

To circumvent this problem one can use the bosonization rules
with the proviso that the fields $g$ and $g^{\dagger}$
must appear pairwised. In other words, an expression like $g^{ij}g^{rs}$
must be written as $-\epsilon^{jl} (g^{\dagger})^{lk} \epsilon^{ki}
g^{rs}$ {\it before} using the rules (\ref{3.7})-(\ref{3.8}). We can
write down the same rule in terms of gauge invariant fermions by stating
that the substitutions
\beq
g^{ij}(z,\bar{z}) \leftrightarrow
\left\{
\begin{array}{ll}
\hat{\psi}^i_2(\bar{z}) \hat{\psi}^{j\dagger}_2(z) &
\mbox{when $g$ occurs as an odd factor} \\
\epsilon^{jl}\hat{\psi}^l_1 (z)
\epsilon^{ik}\hat{\psi}^{k\dagger}_1(\bar{z}) &
\mbox{when $g$ occurs as an even factor}
\end{array}
\right.
\eeq{3.10}
are the appropriate ones when dealing with a multi-point product of
$g$-fields.
In this respect, the literature about
the spinon formulation of $SU(2)_1$ WZW theory \cite{Ludwig} makes
reference to the holomorphic part of the $g$-field only and makes
implicit
use of the relation (\ref{3.9}) any time that a spinon field occurs as
an even factor in a multi-spinon operator product. This gives rise to
two different ``sectors'' in the spinon Fock space: the vacuum sector,
consisting of states created by an even number of spinon operators, and
the spin 1/2 sector, containing states created by an odd number of such
operators.
To conform to the usual notation, we then define a holomorphic field
$\sigma^j(z)$ as:
\beq
\sigma^j(z)=\left\{
\begin{array}{ll}
\hat{\psi}^{j\dagger}_2(z) & \mbox{when acting on the vacuum
sector} \\
\epsilon^{jk} \hat{\psi}^k_1(z) &  \mbox{when acting on the spin 1/2
sector}
\end{array}\right.
\eeq{3.10a}
(The phase for each holomorphic factor is chosen for later convenience;
it is of course irrelevant while considering $g$ correlators).

The $\sigma$-$\sigma$ OPE is now readily evaluated by using
eq.(\ref{3.6}). It is given by
\beq
\sigma^i(z)\sigma^j(w)=(-1)^q \frac{\epsilon^{ij}}{(z-w)^{1/2}}
+ \dots
\eeq{3.10b}
where $q=0$  when this product acts on the vacuum sector and $q=1$
otherwise. As pointed out in \cite{Ludwig}, it is this OPE
what characterizes the $\sigma$'s as spinon fields.
We have then shown that the holomorphic gauge-invariant composite
fermions, as given in the $\sigma^j$ definition, provide a Lagrangian
construction for spinon fields.

In view of the discussion above, the $\sigma^j$ field fulfills the
properties of the bosonic spinon field. Once this identification is
made, one can follow the same procedure described in
\cite{Bernard,Ludwig} to construct the full chiral Hilbert space of the
$SU(2)_1$
CFT from the action of the modes of the $\sigma$ field. For the sake of
conciseness we summarize this construction here.

Due to the square root branch cut in the OPE singularity (\ref{3.10b})
the excitations created by the spinon fields are semionic. This in turn
implies that their mode expansion will involve integer (half-integer)
powers of $z$ when acting on the vaccum (spin $1/2$) sector:
\beq
\sigma^i(z)=\sum_{m=0}^{\infty} z^{m+\frac{q}{2}}
\sigma^i_{-m-\frac{1}{4}-\frac{q}{2}},
\eeq{3.i}
where
\beq
\sigma^i_{-m+\frac{3}{4}-\frac{q}{2}}=\oint\frac{dz}{2\pi i}
z^{-m-\frac{q}{2}}\sigma^i(z).
\eeq{3.i'}

These modes satisfy generalized commutation relations which can be
obtained from (24), giving
\beq
\sum_{l=0}^{\infty} C_l^{(-1/2)}
\left(\sigma^i_{-m-\frac{q+1}{2}-l+\frac{3}{4}}
\sigma^j_{-n-\frac{q}{2}+l+\frac{3}{4}}-
\sigma^j_{-n-\frac{q+1}{2}-l+\frac{3}{4}}
\sigma^i_{-m-\frac{q}{2}+l+\frac{3}{4}}\right)
= (-1)^q \epsilon^{ij} \delta_{m+n+q-1,0}
\eeq{3.j}
where $C_l^{(-1/2)}$ are the coefficients of the Taylor expansion of
$(1-z)^{-1/2}$ \cite{FZ}.

The chiral part of the Fock space of the $SU(2)_1$ conformal field
theory can be constructed in terms of these modes. As stated above, this
space can be classified into multiplets corresponding to the
irreducible representations of the Yangian algebra $Y(sl_2)$.
Each multiplet is
constructed from a highest weight state by the action of the Yangian
generators.
The highest weight states can be constructed in
terms of the spinon modes in the following way:
one has to take linear combinations of $M$-spinon states
\beq
\sigma^{1}_{-n_1-1/4} \cdots \sigma^{1}_{-n_M-1/4}|0\rangle
\eeq{3.k}
that are annihilated by the raising operators $Q^+_0$ and $Q^+_1$.
These states correspond to the ``fully polarized $M$-spinon sates'' of
Haldane
\cite{Hal-spinon}. The defining level $0$ and $1$ generators of $Y(sl_2)$
can be written in terms of the modes of the Kac-Moody currents
\cite{Hal-al}
as
\beqn
Q_0^a &=& J_0^a, \nonumber\\
Q_1^a &=& \frac{1}{2}f^a_{bc}\sum_{m=0}^{\infty} :J_{-m}^bJ_m^c:.
\eeqn{3.k'}

The irreducible Yangian multiplet corresponding to a given highest weight
state is then constructed by repeatedly applying the generators $Q_0^-$
and
$Q_1^-$.
Finally, the union of all of these multiplets forms a basis of the
full Hilbert space.

\vspace{5mm}

\noindent {\it iv) Holomorphic factorization of other primaries in
$SU(N)_1$ and their relation with $SU(N)_1$ spinon fields}

\vspace{5mm}

In section {\it ii)} we presented the holomorphic factorization of the
primary field $g$ in the $SU(N)_1$ WZW theory. In the general
case $N>2$ the theory contains $N$ primary fields, each one
corresponding to an integrable representation, which can
be constructed as
suitable symmetrized products of fields in the fundamental
representation. We will now show how the holomorphic factors of all
these primaries can be
explicitly constructed in terms of the gauge invariant fermions (8).

The integrable representations for the
$SU(N)_1$ case correspond to Young diagrams with one column and at most
$N-1$ rows. The primary fields in these representations, $\Phi_p$, are
then obtained by taking normal-ordered antisymmetric products of
$p$ fields ($p=0,1,\dots N-1$) in the fundamental representation
\cite{NS}.

In terms of the composite fermions (8) the holomorphic parts of the
primaries corresponding to $p=2,\dots ,N-1$ are given by
\beq
\Phi_p^{i_1,i_2,...,i_p}(z)={\cal{A}} \left(:{\hat{\psi}}_1^{i_1}
{\hat{\psi}}_1^{i_2}...{\hat{\psi}}_1^{i_p}:\right) ,
\eeq{3.l}
while in terms of the decoupled fields they can be written as
\beq
\Phi_p^{i_1,i_2,...,i_p}(z)={\cal{A}} \left(:\chi_1^{i_1}
\chi_1^{i_2}...\chi_1^{i_p}:\right):\exp (-p\varphi) : .
\eeq{3.m}

The conformal dimensions of these fields can be evaluated as the sum of
two independent contributions: $p/2$ coming from the $p$ free fermions
and $-p^2/2N$ coming from the vertex operator $:\exp (-p\varphi) :$. The
result is
\beq
h_p=\frac{p}{2}-\frac{p^2}{2N} = \frac{p(N-p)}{2N} .
\eeq{3.n}
These dimensions agree with the conformal dimensions of the $SU(N)_1$
primary
fields (see for instance \cite{KZ}), each one in the representation
$\Lambda_p$ (associated with a Young diagram with $p$ vertical boxes),
given by
\beq
h_{\Lambda_p}=\frac{c_{\Lambda_p}}{c_v+k}
\eeq{3.o}
where $c_v=N$ for $SU(N)$, $k=1$ is the Kac-Moody level and
$c_{\Lambda_p}=\frac{p}{2N}
(N+1)(N-p)$ is the Casimir of the representation $\Lambda_p$.

The Yangian structure is also present in the $N>2$ case \cite{Sch}, but the
spinon construction has not yet been presented.

\vspace{5mm}

\noindent{\it v) Conclusions}

\vspace{5mm}

We have proved that the physical operators in the fermionic coset
realization of $SU(N)_1$ CFT are automatically holomorphic or
antiholomorphic. The standard bosonization rules, when written in terms
of gauge invariant fermions, provide an explicit holomorphic
factorization of WZW primary fields. Moreover, we have shown how to
express the holomorphic part of any primary in the $SU(N)_1$ theory.

As expected, the holomorphic part
of the spin 1/2 primary field in the $SU(2)_1$ case fulfills
the characterization of the so called spinon field. As a consequence,
the fermionic coset model $U(2)/U(1)$ provides a Lagrangian formulation
for spinons, which are known to behave as semionic $1+1$
dimensional quasiparticles.

The present description of the spinon representation of $SU(2)$ CFT in
terms of composite fermions in a constrained fermionic formulation seems to
be, in our opinion, a natural framework to investigate some open
questions.
In the same line, the fundamental r\^ole played by composite gauge
invariant fermions has already shown up in the coset formulation of the
Ising model \cite{CR1} (see also \cite{CM}), where they were used to build up 
the order and
disorder operators. We expect that this guideline should allow for the
holomorphic factorization of some primaries in more general coset
models.

It will be also interesting to see whether an underlying
structure similar to
the Yangian appearing in $SU(N)_1$ is also present in other CFTs, such as
higher level $SU(N)$ CFT or coset models. In fact, the holomorphic
factorization of the WZW primaries in the $SU(N)_k$ case ($k>1$) in
terms of gauge invariant fermions can be
envisaged following the same lines as those presented here. The
connection with the spinon basis constructed in Ref. \cite{Ludwig2}
will be discussed elsewhere \cite{CR}.

\vspace{.5cm}

{\it  Acknowledgements}: we would like to thank J. Fuchs and K.
Schoutens for useful discussions and specially E.F.Moreno for his
helpful suggestions and valuable comments. We also thank F. A.
Schaposnik for comments on the preliminar manuscript.

\newpage

{\it Note added}: after this paper was finished, the spinon construction has 
been extended to the $N>2$ case \cite{BS96}. The complete (chiral) Fock space 
has been constructed there by acting on the vacuum with the modes of the 
holomorphic factor of the primary field transforming in the $\bar N$ 
representation of $SU(N)$ (in their notation, $\phi^{\alpha}(z)$, 
$\alpha = 1,2,\dots ,N$).
As stressed in \cite{BS96}, the statistics of these fields is encoded in their
OPE. In their notation, it is given by
\beq
\phi(z)\phi(w)=\frac{\phi ' (w)}{(z-w)^{1/N}}+\dots
\eeq{4.13}
where $\phi'$ is the primary field in the representation $\Lambda_{N-2}$ 
of $SU(N)$. 

This construction can of course be made out from the fields we present in 
section $iv)$. In particular, it is simpler in our notation to work with the 
representation $N$ instead of $\bar N$, and define the spinon creation 
operators from the modes of $\Phi_{p=1}$. Using our expression (26) for this 
field, we can easily evaluate the OPE in terms of the decoupled fields:
\beq
\Phi_1^i(z) \Phi_1^j(w)=\frac{1}{(z-w)^{1/N}}{\cal A}
(:\chi_1^i(w)\chi_1^j(w):)
:\exp (-2\varphi(w)):+\dots  = \frac{1}{(z-w)^{1/N}} \Phi_2^{ij}(w) + \dots
\eeq{toma}
The agreement between (\ref{4.13}) and (\ref{toma}) shows that the gauge 
invariant fermions (\ref{3.2}) provide an explicit Lagrangian realization of 
the spinon fields not only for $N=2$ but for arbitrary $N$.


\newpage

\noindent {\it Appendix: Yangian structure of the Fock space of
$SU(N)_1$ conformal field theories}

\vspace{0.5cm}

We review here the main aspects of a novel representation of
the spectrum of the chiral part of $SU(N)_1$ CFTs, first introduced
by Haldane et al.\cite{Hal-spinon,Hal-al} and further developed in
\cite{Bernard,Ludwig,Sch,Ludwig2}.
It was originally presented in \cite{Hal-spinon} as an extrapolation of
the exact solution
of long range interaction 1D spin chains \cite{Hal-old,Shastry} and the
fact that
this chain is described in the low energy limit by the level one $SU(2)$
conformal field theory \cite{Affleck}.

In the $SU(2)_1$ case the new representation, known as spinon
representation, allows for a quasi-particle
interpretation: in this sense, each state corresponds to a given number
of spin one-half, neutral, semionic, free quasi-particles (spinons).

The quantum numbers that characterize the representation correspond to the
behavior
of the states under the action of the generators $Q_n^a$ of an
infinite algebra, known as
Yangian algebra \cite{Drinfel'd} (where $n=0,1,\dots$ and $a$ is a
Lie algebra index).
The Yangian $Y(\hat{g})$ associated to a Lie algebra
$\hat{g}$ is a Hopf algebra that is neither commutative nor cocommutative,
being a simple
non-trivial example of a quantum group.

The Yangian was defined in \cite{Drinfel'd} by starting with the generators
$Q_0^a$ and $Q_1^a$,
which for $\hat{g}=sl_N$ (the Lie algebra of $SU(N)$) are defined
by the relations
\beqn
{[}Q_0^a, Q_0^b{]} &=& f^{abc}Q_0^c,\nonumber \\
{[}Q_0^a, Q_1^b{]} &=& f^{abc}Q_1^c,\nonumber \\
{[}Q_1^a,{[}Q_1^b, Q_0^c{]}{]}+(cyclic\: in\:
a,b,c) &=& A^{abcdef}\{Q_0^d,Q_0^e,Q_0^f\}
\nonumber \\
{[}{[}Q_1^a, Q_1^b{]},{[}Q_0^c, Q_1^d{]}{]}+{[}{[}Q_1^c,
Q_1^d{]},{[}Q_0^a, Q_1^b{]}{]} &=&
(A^{abpqrs}f^{cdp}+A^{cdpqrs}f^{abp})\{Q_0^q,Q_0^r,Q_1^s\},
\eeqn{Q1}
where  $A^{abpdef}=\frac{1}{4} f^{adp}f^{beq}f^{cfr}f^{pqr}$ and
$\{...,...,...\}$ denotes a completely symmetrized product.
Higher level generators are recursively defined for $n\ge 2$ by
\beq
c_vQ_n^a=f^{abc}{[}Q_1^c,Q_{n-1}^b{]},
\eeq{Q2}
where $c_v$ is the Casimir in the adjoint representation.

A central property of this algebra is the comultiplication rule
that can be viewed as a generalization of angular momentum addition:
\beq
\begin{array}{l}
\Delta_{\pm}(Q_0^a)=Q_0^a \otimes 1 + 1 \otimes Q_0^a\\
\Delta_{\pm}(Q_1^a)=Q_1^a \otimes
1 + 1 \otimes Q_1^a \pm\frac{1}{2}
f^{abc} Q_0^b \otimes Q_0^c  .
\end{array}
\eeq{Qn}

Being the Yangian a symmetry of the $SU(N)_1$ CFT the whole Hilbert
space can be decomposed into a sum of multiplets corresponding to its
irreducible representations.
Each multiplet is characterized by a highest weight state defined by
\beq
Q_n^{\alpha}|\psi\rangle =0
\eeq{Q3}
where $\alpha$ corresponds to a positive root in the Cartan basis.
Its vectors are eigenstates of
$Q_n^i$, where $i$ corresponds to Cartan subalgebra generators.
Each highest weight state
is characterized by a finite set of non-decreasing integers
$\{n_i\}_{i=1,\dots
,p}$. In the $SU(2)_1$ quasi-particle language each one of these
integers represents one spinon in the $n_i$-th orbital.

\end{document}